\documentstyle[numreferences]{kapproc}

\def\openone{\leavevmode\hbox{\small1\kern-3.8pt\normalsize1}}%

\def\overcirc#1{\protect
\setbox0=\hbox{$\displaystyle #1$}%
\setbox1=\hbox{$\scriptstyle \circ$}%
\setbox2=\hbox{}\ht2=\ht0 \dp2=\dp0 %
\ifdim\wd0>\wd1 %
\setbox1=\hbox to\wd0{\hss\box1\hss}%
\mathord{\rlap{\raise1.2\ht0\box1}\box0}%
\else   %
\setbox1=\hbox to.9\wd1{\hss\box1\hss}%
\setbox0=\hbox to\wd1{\hss$\displaystyle\relax#1$\hss}%
\mathord{\rlap{\copy0}\raise1.2\ht0\box1}%
\fi
\mathord{\box2}}

\runningtitle{MAJORANA MODELS}
\runningauthor{Dvoeglazov}

\begin{opening}

\title{Majorana -- like models in the physics of neutral
particles\thanks{Reported at
the XXX Escuela Latino Americana de F\'{\i}sica, M\'exico city   (July 17-
August   4, 1995); the Int. Conference on the Theory of the Electron,
ICTE'95, Cuautitl\'an, M\'exico (Sept. 27-29, 1995) and  at  the  XXXVIII
Congreso Nacional de la SMF, Zacatecas, M\'exico (Oct. 16-20, 1995).}}

\author{Valeri V. Dvoeglazov\thanks{On leave of absence from
{\it Dept. Theor. \& Nucl. Phys., Saratov State University,
Astrakhanskaya ul., 83, Saratov\, RUSSIA.}  Internet address:
dvoeglazov@main1.jinr.dubna.su}}

\institute{Escuela de F\'{\i}sica, Universidad Aut\'onoma de Zacatecas \\
Antonio Doval\'{\i} Jaime\, s/n, Zacatecas 98068, ZAC., M\'exico\\
Internet address: VALERI@CANTERA.REDUAZ.MX}

\end{opening}

\begin{document}
\begin{center} Submitted 21 October  1995 \end{center}

\begin{abstract}
Due to the standard   electroweak model   we have become accustomed
to think about a neutrino $\nu$ and its antineutrino $\overline \nu$ as
distinct particles. However, it has long been recognized that the apparent
distinction between them may be only an illusion.
Implying these words of   Prof.  B.  Kayser we give an alternative insight
in the physics of neutral particles (neutrino and photon).   The proposed
formalism which is based on the  Majorana ideas could also be useful for
deeper understanding of the nature of   electron.

\end{abstract}


\bigskip
\bigskip

The   standard electroweak model tells us that neutrino can be only
left-handed and   antineutrino can be only right-handed.   Neutrino and
its antineutrino are different particles in the framework of conventional
approaches. However, the standard  model, which is built on the base of
the Dirac   construct   for fermions and of
the gauge principle, is not able to explain us, why the mirror image of
the $\beta$- decay does not occur in Nature, where are `missing'
right-handed neutrino and left-handed antineutrino?  In this talk I try to
use a viewpoint based on the Majorana ideas in order to understand
mathematical origins of these facts.  First of all, let me discuss the
present situation in the physics of neutral particles (as well as in
Physics itself).  At the moment we have:

\begin{itemize}
\item
the solar neutrino puzzle;

\item
the negative mass squared problem ({\it cf.} with the old ITEP
result (1987));

\item
the atmospheric neutrino anomaly;

\item
speculations on the possibility of the neutrinoless
double $\beta$-decay;

\item
the experimental evidence for tensor coupling in the decay
$\pi^- \rightarrow e^- +\overline \nu_e +\gamma$
(as well as in  the decays of $K$ mesons);

\item
the dark matter problem;

\item
the problem of $\gamma$-ray bursts;

\item
candidate events for neutrino oscillations (LANL, April 1995),
what, according to present-of-day ideas, could lead to the conclusion
of  existence of the neutrino mass and, probably, of
the fourth generation;

\item
in addition, the spin crisis in QCD.
\end{itemize}

Furthermore, there are several theoretical puzzles in
basic structures of  the quantum field theory (QFT):
\begin{itemize}

\item
In classical physics antisymmetric tensor field is transversal;
on the other hand it was proved
that `after quantization' the antisymmetric tensor field
is longitudinal.   Does this fact signify that
we must abandon the Correspondence Principle?\ldots

\item
The problem of the indefinite metric in QFT. Nobody appears to understand
its physical sense. Moreover, from a viewpoint of  mathematics it is rather
obscure construct.

\item
Finally, the renormalization idea, which   ``would be
sensible only if it was applied with finite renormalization factors, not
infinite ones (one is not allowed to neglect and
[to subtract] infinitely large quantities)". These words are not of mine
but the words of Prof. Dirac presented in his last
lectures~[1a,p.4-5].
\end{itemize}

While I am not going to answer all these questions in the present   talk,
but  quite a number of  these problems seems to me to be overwhelming
if compare with
only three black spots (black-body radiation, photoeffect
and $\alpha$- particle scattering)
on the cloudless sky of the classical physics in the end of
the nineteenth century.

Various models have been proposed for explanation of the present situation.
I list some of them:
\begin{itemize}

\item
Non-zero electric charge of neutrino~\cite{Ignatiev}.
These models are based on the realization of the old Einstein's
idea of the electric charge dequantization.

\item
Existence of the mirror matter (as   a particular case, of the mirror
photon with electric charge and/or mass), ref.~\cite{Foot,Giveon}.

\item
Neutrino theory of light ({\it e.g.}, ref.~\cite{Bandy}). ``\ldots In view
of the neutrino theory of light, photons are likely to interact weakly also,
apart from the usual electromagnetic interactions\ldots This assumed
photon-neutrino weak
interaction, if it exists, will have important bearing on astrophysics\ldots"

\item
Tachyonic neutrinos, ref.~\cite{tachyon}.

\item
Introduction of the Evans-Vigier longitudinal $B(3)$ field
of electromagnetism, ref.~\cite{Evans}. Let me remind that many physicists
(including Dirac~[1b,p.32]) have risen the problem of longitudinal
modes, {\it e.g.},~\cite{Pappas,Graneau}. In my opinion, recent preprints
and papers~\cite{PASH}-\cite{Chubykalo}  proved that the problem exists.

\item
The use of other representations of the extended Lorentz group (namely,
the doubled representations~\cite{Wigner2}), see, {\it e.g.},
ref.~\cite{DVA1}.

\end{itemize}

Several of these models are certainly exotic if not say   `crazy'.
But, I hope, you remember
the known saying of the great physicist of the twentieth
century: ``The idea is
crazy. Is it {\it sufficiently} crazy to be correct?"

Our aim with this talk (and with the recent series of our papers) is to
understand the Majorana ideas~\cite{Majorana} of constructing the theory
of neutral particles and to propose models based on these principles
with possible application to neutrino physics.  The work in this direction
has been started in~\cite{DVA2}-\cite{DVO1}.  Apart from these papers
I would also like to mention several papers of Barut and Ziino which deal
with similar matters~\cite{Ziino,Barut}.

First of all, let me touch the Dirac case:
\begin{equation}
\left [ i\gamma^\mu \partial_\mu -m \openone \right ]
\Psi (x) = 0\quad.\label{de}
\end{equation}
Everybody knows the physical sense of the
spinorial basis (in the standard representation of $\gamma$ matrices)
\begin{eqnarray}
u^{(1)} (\overcirc{p}^\mu) &=& \sqrt{m}\pmatrix{1\cr 0\cr 0\cr 0\cr},\quad
u^{(2)} (\overcirc{p}^\mu) = \sqrt{m}\pmatrix{0\cr 1\cr 0\cr 0\cr},\quad
\label{sbde1}\\
v^{(1)} (\overcirc{p}^\mu) &=& \sqrt{m}\pmatrix{0\cr 0\cr 1\cr 0\cr},\quad
v^{(2)} (\overcirc{p}^\mu) = \sqrt{m}\pmatrix{0\cr 0\cr 0\cr
1\cr}\label{sbde2}
\end{eqnarray}
($\overcirc{p}^\mu$  should be referred
to the frame with the momentum ${\bf  p}\rightarrow {\bf 0}$). The
equation (\ref{de}) describes eigenstates of the Charge Operator.  One can
attach particle-antiparticle interpretation of bispinors
(\ref{sbde1},\ref{sbde2}).  But, if apply the Majorana {\it ansatz} (in
the coordinate representation; signs $\pm$ are referred to left and right
projections $\Psi_\pm = {1\over \sqrt{2}} (1\pm \gamma_5)\Psi$):
\begin{equation}
\Psi_- (x) = {\cal C}_{[1/2]} \Psi_+^\ast (x)
\end{equation}
one can obtain neutral
particles. Why?  This question has been analyzed in ref.~\cite{MLC}. A
review of topics connected with this interpretation of the Majorana ideas
could be found in ref.~\cite{Mannheim}.

I am going to consider models based on the following  very general postulates:
\begin{itemize}
\item
For arbitrary $j$ the right $(j,0)$ and the left $(0,j)$ handed spinors
transform according to the Wigner's rules~\cite{Wigner,Wigner2}:
\begin{eqnarray}
\phi_{_R} (p^\mu)\, &=& \,\Lambda_{_R} (p^\mu \leftarrow
\overcirc{p}^\mu)\,\phi_{_R}
(\overcirc{p}^\mu) \, = \, \exp (+\,{\bf J} \cdot \mbox{\boldmath
$\varphi$}) \,\phi_{_R} (\overcirc{p}^\mu)\quad,\label{w1}\\ \phi_{_L}
(p^\mu)\, &=&\, \Lambda_{_L} (p^\mu \leftarrow
\overcirc{p}^\mu)\,\phi_{_L} (\overcirc{p}^\mu) \, = \, \exp (-\,{\bf J}
\cdot \mbox{\boldmath $\varphi$})\,\phi_{_L}
(\overcirc{p}^\mu)\quad.\label{w2} \end{eqnarray} $\Lambda_{_{R,L}}$ are
the matrices of Lorentz boosts; ${\bf J}$ are the spin matrices for spin
$j$; $\mbox{\boldmath $\varphi$}$ are parameters of the given boost.   If
restrict ourselves by the case of bradyons they are defined, {\it e.~g.},
refs.~\cite{Ryder,DVA1}, by means of:
\begin{equation}\label{boost}
\cosh (\varphi) =\gamma = \frac{1}{\sqrt{1-v^2}} = \frac{E}{m},\quad
\sinh (\varphi) = v\gamma = \frac{\vert {\bf p}\vert}{m},\quad \widehat
{\mbox{\boldmath $\varphi$}} = {\bf n} = \frac{{\bf p}}{\vert {\bf
p}\vert}\quad.
\end{equation}

\item
$\phi_{_L}$ and $\phi_{_R}$ are the eigenspinors of the helicity
operator $({\bf J}\cdot {\bf n})$:
\begin{equation}
({\bf J}\cdot {\bf n})\,\phi_{_{R,L}} (p^\mu) \,=\, h \,\phi_{_{R,L}} (p^\mu)
\end{equation}
($h = -j, -j+1,\ldots j$   is the helicity quantum number).

\item
The relativistic dispersion relations
$E = \pm \sqrt{{\bf p}^2 +m^2}$ are hold for observed particle states.

\end{itemize}

Apart from these mathematical postulates the cornerstone of
relativistic theories appears to be the Ryder-Burgard (RB) relation.
Ryder~\cite{Ryder} writes:  ``When a particle is at rest, one cannot
define its spin as either left- or right-handed, so $\phi_{_R} (0) =\phi_{_L}
(0)$." In order to include negative-energy solutions it is necessary to
take into account the possibility of opposite sign in
the relation for right- and left- spinors~\cite{DVA1}.
On the base of the   RB relation
and the postulates which is above (the Wigner rules for the Lorentz
transformations) the Dirac equation follows immediately,
ref.~[20b].  ``Refer to Eqs. (\ref{w1}) and (\ref{w2}) and set
${\bf J} = \mbox{\boldmath $\sigma$}/2$. Next, note that spinors [implied
by the arguments based on parity symmetry and that Lorentz group is
essentially $SU_R (2) \otimes SU_L (2)$]
\begin{equation}
\psi (p^\mu) =
\pmatrix{\phi_{_R} (p^\mu)\cr \phi_{_L} (p^\mu)}
\end{equation}
turn out to be
of crucial significance in constructing a field $\Psi (x)$ that describes
eigenstates of the Charge operator, $Q$, if
\begin{equation}\label{rb}
\phi_{_R} (\overcirc{p}^\mu)
=\pm \phi_{_L} (\overcirc{p}^\mu)
\end{equation}
(otherwise physical eigenstates are no longer
charge eigenstates). We call [this relation], the ``Ryder-Burgard
relation"... Next couple the RB relation with Eqs.
(\ref{w1}) and (\ref{w2}) to obtain
\begin{eqnarray}
\pmatrix{\mp m \,\openone & p_0 + \mbox{\boldmath $\sigma$}\cdot {\bf p}\cr
p_0 - \mbox{\boldmath $\sigma$}\cdot {\bf p} & \mp m \,\openone\cr} \psi
(p^\mu)\, =\, 0\quad.
\end{eqnarray}
[Above we have used the property
\begin{equation}
\left [\Lambda_{_{L,R}} (p^\mu
\leftarrow \overcirc{p}^\mu)\right ]^{-1} =
\left [\Lambda_{_{R,L}} (p^\mu \leftarrow \overcirc{p}^\mu)\right ]^\dagger
\end{equation}
and that both ${\bf J}$ and $\Lambda_{_{R,L}}$ are Hermitian
for the finite, {\it e.g.},  $(1/2,0)\oplus (0,1/2)$ representation
of the Lorentz group].
Introducing $\Psi (x) \equiv \psi (p^\mu)   \exp (\mp ip\cdot x)$
and letting $p_\mu \rightarrow i\partial_\mu$, the above equation
becomes: $(i\gamma^\mu \partial_\mu - m \,\openone) \Psi (x) = 0$.
This is the Dirac equation for spin-1/2 particles with
$\gamma^\mu$ in the Weil/Chiral representation. Similarly,
one can obtain wave equations and thus a complete kinematic structure
and the associated dynamical consequences for other Dirac-like
$(j,0)\oplus (0,j)$ spinors $\psi (p^\mu)$ and quantum fields
$\Psi (x)$."

Let me now consider  generalized cases.
\begin{enumerate}

\item

\begin{equation}
\phi_{_R}^\pm (\overcirc{p}^\mu) = {\cal A} \phi_{_L}^\pm
(\overcirc{p}^\mu)\quad.
\end{equation}
${\cal A}$ is the matrix of arbitrary linear transformation. Expanding it
in the complete set of $\sigma_i$ matrices (${\cal A} \equiv   \openone c_1^0
+\mbox{\boldmath $\sigma$} \cdot  {\bf c}_1 $)
one can easily obtain:
\begin{equation}
\phi_{_R}^\pm
(\overcirc{p}^\mu) = e^{i\alpha_{\pm}} \phi_{_L}^\pm (\overcirc{p}^\mu)\quad.
\end{equation}
We have used the second postulate.
In spite of the fact that, in general, ${\bf c}_1  \not \parallel {\bf p}$
this is possible for $\phi_{_{R,L}} (\overcirc{p}^\mu)$ spinors as was
explained in ref.~\cite[p.93]{Sakurai}.

The way of  deriving the equation is the same to the above:
\begin{eqnarray}
\phi_{_R}^\pm (p^\mu) \,&=&\, \Lambda_{_R} (p^\mu \leftarrow
\overcirc{p}^\mu)\,
\phi_{_R}^\pm (\overcirc{p}^\mu) = \,e^{i\alpha_{\pm}}
\,\Lambda_{_R}   (p^\mu \leftarrow \overcirc{p}^\mu)
\,\phi_{_L}^\pm (\overcirc{p}^\mu) =\nonumber  \\
&=&\, e^{i\alpha_{\pm}}\,\Lambda_{_R} (p^\mu
\leftarrow \overcirc{p}^\mu) \,\Lambda_{_L}^{-1} (p^\mu \leftarrow
\overcirc{p}^\mu)\,\phi_{_L}^\pm (p^\mu)\quad, \label{eq1} \\
&&\nonumber\\
\phi_{_L}^\pm (p^\mu) \,&=&\, \Lambda_{_L} (p^\mu \leftarrow \overcirc{p}^\mu)
\,\phi_{_L}^\pm (\overcirc{p}^\mu) \, = \, e^{-i\alpha_{\pm}}\,
\Lambda_{_L} (p^\mu \leftarrow \overcirc{p}^\mu) \, \phi_{_R}^\pm
(\overcirc{p}^\mu) =\nonumber\\
&=& \,e^{-i\alpha_{\pm}} \,\Lambda_{_L} (p^\mu
\leftarrow \overcirc{p}^\mu) \,\Lambda_{_R}^{-1} (p^\mu \leftarrow
\overcirc{p}^\mu) \,\phi_{_R}^\pm (p^\mu)\quad.\label{eq2}
\end{eqnarray}
Using definitions of the Lorentz boost (\ref{w1}-\ref{boost})
one can re-write the equations (\ref{eq1},\ref{eq2}) in  matrix form
(provided that $m\neq 0$):
\begin{eqnarray}
\pmatrix{-m e^{-i\alpha_{\pm}}& p_0 +(\mbox{\boldmath $\sigma$}\cdot{\bf
p})\cr p_0 - (\mbox{\boldmath $\sigma$}\cdot{\bf p}) & -m
e^{i\alpha_{\pm}}\cr} \pmatrix{\phi_{_R} (p^\mu) \cr \phi_{_L} (p^\mu)} =
0\quad,
\end{eqnarray}
or
\begin{equation}
\left ( \hat p -m{\cal T}
\right ) \psi (p^\mu) = 0\quad,
\end{equation}
with
\begin{eqnarray}
{\cal
T} = \pmatrix{e^{-i\alpha_{\pm}}& 0\cr 0 & e^{i\alpha_{\pm}}\cr}\quad.
\end{eqnarray}
Particular cases are:
\begin{eqnarray}
\alpha_\pm &=& 0, 2\pi \quad : \quad (\hat p - m)\psi (p^\mu)
=0\quad, \label{DE1}\\
\alpha_\pm &=& \pm\,\, \pi \quad\,\,\, : \quad (\hat p + m)\psi
(p^\mu)=0\quad, \label{DE2}\\
\alpha_\pm &=& +\pi/2 \quad : \quad (\hat p
+ im\gamma_5)\psi (p^\mu) =0\quad,\label{DDE1}\\
\alpha_\pm &=& -\pi/2
\quad :  \quad (\hat p - im\gamma_5)\psi (p^\mu) =0 \quad.  \label{DDE2}
\end{eqnarray}
Equations (\ref{DE1},\ref{DE2}) are
the well-known Dirac equations for positive- and negative-energy bispinors
in the momentum space.  Equations of the type (\ref{DDE1},\ref{DDE2})
had also been discussed in the old literature, {\it e.~g.},
ref.~\cite{Sokolik}.  They have been named as the Dirac equations for
4-spinors of the second kind~\cite{Cartan,Gelfand,DVA3}.  Their possible
relevance to describing neutrino had been mentioned in the cited papers.

\item
Another generalization deals with the  RB relation in the other form,
which is not equivalent to the first one (let me remind that
complex conjugation is not a linear operator):
\begin{equation}
\phi_{_R} (\overcirc{p}^\mu) ={\cal B} \phi_{_L}^\ast
(\overcirc{p}^\mu)\quad.\label{rb2}
\end{equation}
In this case  we use expansion in the different complete set of
$\sigma_i$ matrices (let me remind a mathematical theorem that
after multiplying each of matrices, which form a complete set, by  a
non-singular matrix the property to be the complete set is hold).
We come to the equation (\ref{rb2}) re-written in convenient form:
\begin{eqnarray}
\lefteqn{\phi_{_R}^\pm (\overcirc{p}^\mu) = {\cal B} [\phi_{_L}^\pm
(\overcirc{p}^\mu)]^* = \left [ c^0_2 \, \sigma_2 + (\mbox{\boldmath
$\sigma$}\cdot {\bf c}_2) \sigma_2\right ] [\phi_{_L}^\pm
(\overcirc{p}^\mu)]^* =\nonumber} \label{rbc} \\
&=& \left [i\,c^0_2 \,
\Theta_{[1/2]} \mp i\, (\vert {\Re}e\, {\bf c}_2\vert + i \, \vert
{\Im}m\, {\bf c}_2 \vert ) \,\Theta_{[1/2]}\right ] [\phi_{_L}^\pm
(\overcirc{p}^\mu)]^* \nonumber\\
&=& i\,e^{i\beta_{\mp}} \Theta_{[1/2]}\,
[\phi_{_L}^\pm (\overcirc{p}^\mu)]^*
\end{eqnarray}
and, hence, to the
inverse one
\begin{equation}
\phi_{_L}^\pm (\overcirc{p}^\mu) = -
i\,e^{i\beta_{\mp}} \Theta_{[1/2]} [\phi_{_R}^\pm
(\overcirc{p}^\mu)]^*\quad.\label{rbinv}
\end{equation}
We have used above
that $\sigma_2$ matrix is connected with the Wigner operator
$\Theta_{[1/2]}=-i\sigma_2$
and the property of the Wigner operator for any spin
$\Theta_{[j]} {\bf J}\Theta_{[j]}^{-1} = -{\bf J}^*$.
So, if $\phi_{_{L,R}}$ is an eigenstate of the helicity
operator, then $\Theta_{[j]}\phi_{_{L,R}}^*$ is the eigenstate with
the opposite helicity quantum number:
\begin{equation}
({\bf J}\cdot {\bf n}) \,\Theta_{[j]} \left [ \phi_{_{L,R}}^h (p^\mu)\right ]^*
= -\, h \,\Theta_{[j]} \left [\phi_{_{L,R}}^h (p^\mu)\right
]^*\quad.\label{wi}
\end{equation}
Therefore, from Eqs.
(\ref{rbc},\ref{rbinv}) we have
\begin{eqnarray}
\phi_{_R}^\pm (p^\mu) &=& +\,i\,e^{i\beta_{\mp}}\, \Lambda_{_R}
(p^\mu \leftarrow
\overcirc{p}^\mu)\, \Theta_{[1/2]}\, [\Lambda_{_L}^{-1}
(p^\mu \leftarrow \overcirc{p}^\mu)]^* \,
[\phi_{_L}^\pm (p^\mu)]^*\quad,\label{cc1}\\
\phi_{_L}^\pm (p^\mu) &=& -\,i\,e^{i\beta_{\mp}} \,\Lambda_{_L}
(p^\mu \leftarrow
\overcirc{p}^\mu) \,\Theta_{[1/2]}\, [\Lambda_{_R}^{-1}
(p^\mu \leftarrow \overcirc{p}^\mu)]^* \,
[\phi_{_R}^\pm (p^\mu)]^*\quad.\label{cc2}
\end{eqnarray}
Using the mentioned property of the Wigner operator
we transform   Eqs. (\ref{cc1},\ref{cc2}) to
\begin{eqnarray}
\phi_{_R}^\pm (p^\mu) &=& +\,ie^{i\beta_{\mp}} \Theta_{[1/2]}
[\phi_{_L}^\pm (p^\mu)]^*\quad,\\
\phi_{_L}^\pm (p^\mu) &=& -\,ie^{i\beta_{\mp}} \Theta_{[1/2]}
[\phi_{_R}^\pm (p^\mu)]^*\quad.
\end{eqnarray}
Finally, in  matrix form one has
\begin{equation}
\pmatrix{\phi_{_R} (p^\mu)\cr \phi_{_L} (p^\mu)} =
e^{i\beta_{\mp}} \pmatrix{0& i\Theta_{[1/2]}\cr
-i\Theta_{[1/2]} &0\cr}\pmatrix{\phi_{_R}^* (p^\mu)\cr
\phi_{_L}^* (p^\mu)} = S^c_{[1/2]} \pmatrix{\phi_{_R} (p^\mu)\cr
\phi_{_L} (p^\mu)}\, , \label{sc}
\end{equation}
with $S^c_{[1/2]}$ being the operator of charge conjugation in
the $(1/2,0)\oplus (0,1/2)$ representation space, {\it
e.g.}, ref.~\cite{Ramond}.  We obtain, in fact,  conditions of
self/anti-self charge conjugacy:
\begin{equation}
\psi (p^\mu) = \pm \psi^c (p^\mu)\quad.
\end{equation}
Thus, depending on relations between
left- and right-handed spinors (as a matter of fact, depending on the
choice of the spinorial basis) we  can describe physical excitations of
the very different physical nature.

\item
The most general form of the RB relation is:
\begin{equation}
\phi_{_R} (\overcirc{p}^\mu) = {\cal A} \phi_{_L} (\overcirc{p}^\mu) +
{\cal B} \phi_{_L}^* (\overcirc{p}^\mu)\quad,
\end{equation}
The generalized form of the equation in the $(1/2,0)\oplus (0,1/2)$
representation space is then:
\begin{equation}
\left [a \,{\hat p \over m} + b \,{\cal T} \,S^c_{[1/2]}
- {\cal T}\right ]\psi (p^\mu) = 0\quad,\quad a^2 +b^2 =1\quad.
\end{equation}
Using  computer algebra systems, {\it e.g.}, MATEMATICA 2.2 it is easy
to check that the equation has correct relativistic dispersion relations
(see the third item of the set of postulates).

\end{enumerate}

\smallskip

The following part of my talk  is concerned with the `old-fashioned'
formalism proposed by Professor S. Weinberg long ago, namely, the
$2(2j+1)$ formalism~\cite{Joos}-\cite{Tucker}.  In spite of some
antiquity of this formalism,  in our opinion, it does not deserve `to be
retired'.  The equation for a $j=1$ case, proposed in the sixties, is:
\begin{equation}
\left [ \gamma_{\mu\nu} p^\mu p^\nu - m^2 \openone \right ] \Psi
(x) = 0\quad,
\end{equation}
where $\gamma^{\mu\nu}$ are the
Barut-Muzinich-Williams covariantly defined matrices; $p_\mu
=i\partial_\mu$.  However, for the Dirac-like states
the equation has been corrected  recently~\cite{DVA1}:
\begin{equation}
\left [
\gamma_{\mu\nu} \partial^\mu \partial^\nu + \wp_{u,v} m^2 \openone
\right ] \Psi (x) = 0\quad,
\end{equation}
$\wp_{u,v} =\pm 1$. The cited work~\cite{DVA1}
presents itself a  realization of the quantum field theory of the
Bargmann-Wightman-Wigner (BWW) type~\cite{Wigner2} in the $(1,0)\oplus
(0,1)$ representation space.  One can follow the same procedure
which has been applied above in the $(1/2,0)\oplus (0,1/2)$
representation. For a $j=1$ case the first generalization
then yields:
\begin{equation}
\left [ \gamma_{\mu\nu} p^\mu p^\nu   - {\cal T} m^2 \right ] \psi (p^\mu)
= 0\quad.
\end{equation}
The second one is:
\begin{equation}
\psi (p^\mu) = \Gamma_5 S^c_{[1]} \psi (p^\mu)\quad,
\end{equation}
with
\begin{eqnarray}
S^c_{[1]} = \pmatrix{0 &\Theta_{[1]}\cr
-\Theta_{[1]} &0} {\cal K}\quad,\label{gsc}
\end{eqnarray}
and  ${\cal K}$ is the complex conjugation operator.
As opposed to a spin-1/2 case, we have now the condition
of $\Gamma^5 S^c_{[1]}$- conjugacy. So, depending on the choice
of representation space, some subtle mathematical differences
may arise.

I want to do several remarks: 1) In a spin-1 case the wave function
(field operator) $\Psi(x)$ can be re-written in the bivector
form or the antisymmetric tensor form, ref.~\cite{DVO01}; 2) The spin
structure of matrix elements for interaction of two $j=1$ Joos-Weinberg
particles with vector potential is very similar to a spin-1/2 case,
ref.~\cite{DVOW}.  This is certainly an advantage, because this fact
permits us to use many calculations produced for the well-developed
fermion theory; 3) If we use only one field $\Psi(x)$, the $j=1$
Hamiltonian operator is energy-dependent; the Feynman-Dyson propagator is
not equal to the Wick propagator; interaction with the external vector
potential leads to unrenormalizable theory; the theory contains
unexplained tachyonic solutions, {\it etc.} They are, of course,
shortcomings from a viewpoint of modern theory.

Let me consider now models with spinors of the second kind.
Namely, the phase factor takes values $\alpha=\pm {\pi \over 2}$, or
the Ryder-Burgard relation takes a form: $\phi_{_R} (\overcirc{p}^\mu)
=\pm i \phi_{_L} (\overcirc{p}^\mu)$.  Equations in the momentum
representation follow immediately:
\begin{eqnarray}
\left [ i\gamma_5 \hat p - m\right ] \Upsilon_\pm (p^\mu)
&=& 0 \quad,\label{i1}\\
\left [ i\gamma_5 \hat p + m\right ]   {\cal B}_\pm (p^\mu)
&=& 0 \quad.\label{i2}
\end{eqnarray}
Then, equations in the coordinate representation are
\begin{equation}
\left [ \gamma^5 \gamma^\mu\partial_\mu + m\right ] \Psi_{(1)} (x)
= 0 \quad,\quad
\overline \Psi_{(1)} (x)
\left [ \gamma^5 \gamma^\mu\partial_\mu   + m\right ] = 0\quad.
\label{equations}
\end{equation}
They lead to the following theorem, which is easily proved in a
straightforward manner.

\noindent
{\tt Theorem:} One can {\it not} construct the Lagrangian
in terms of independent field variables $\Psi_{(1)}$ and $\overline
\Psi_{(1)}$, that could lead to the Lagrange-Euler equations of the form
(\ref{equations}).

If we still wish to construct the Lagrangian we are forced to introduce
another field satisfying the equations:
\begin{equation}
\left [ \gamma^5 \gamma^\mu\partial_\mu -
m\right ] \Psi_{(2)} (x) = 0 \quad,\quad \overline \Psi_{(2)}  (x) \left [
\gamma^5 \gamma^\mu\partial_\mu   - m\right ] = 0\quad.
\end{equation}
Their characteristic feature is the opposite sign
at the mass term (comparing with $\Psi_{(1)} (x)$) in the coordinate
representation ({\it cf.} also with new models in the $(1,0)\oplus (0,1)$
space).  Spinors satisfying equations
(\ref{i1},\ref{i2}) are not the eigenspinors of the Parity operator.
These  facts hint that we obtain another example of the quantum
field theory discussed in ref.~\cite{Wigner2,DVA1}.  The corresponding
Lagrangian can be written as follows:
\begin{eqnarray}
{\cal L} &=&
{1\over 2} \left [ \overline \Psi_2 \gamma^\mu \gamma^5 \partial_\mu
\Psi_1 + \partial_\mu \overline \Psi_1 \gamma^\mu \gamma^5 \Psi_2 -
\overline \Psi_1 \gamma^\mu \gamma^5 \partial_\mu \Psi_2 - \partial_\mu
\overline \Psi_2 \gamma^\mu \gamma^5 \Psi_1 \right ] - \nonumber\\ &-& m
\left [ \overline \Psi_1 \Psi_2 + \overline \Psi_2 \Psi_1 \right
]\quad.\label{lag}
\end{eqnarray}
Physical consequences of this model are
the following:
\begin{itemize}

\item
Depending on relations between creation and annihilation operators
of $\Psi_{(1)}$ and $\Psi_{(2)}$ we can describe  charged particles
in the $(1/2,0)\oplus (0,1/2)$ representation space, but
also one can obtain neutral particles.

\item
Formalism admits (in particular cases) the use of either
commutation or anticommutation relations. One can describe bosons
in the $(1/2,0)\oplus (0,1/2)$ representation space.

\item
There is a puzzled physical `excitation'
with $E\equiv 0$, $Q \equiv 0$ and $( W \cdot n) \equiv 0$.

\item
Transitions $\Psi_{(1)} \leftrightarrow \Psi_{(2)}$ are possible. In order
to calculate contributions to self-energies and/or vertex functions
development of the Feynman diagram technique
(or other methods of higher order calculations) is required.

\item
The role of the Feynman-Dyson propagators is similar to the Dirac theory:
to propagate positive-frequency solutions toward positive times and
the negative-frequency ones, backward in time, {\it i.e.}, it is compatible
with the Feynman-St\"uckelberg scheme. But, the Feynman-Dyson propagators
are not the Wick propagators ({\it cf.} with a $j=1$ case).

\item
As a result of analysis of this model the question arises: if start from
the conventional Dirac Lagrangian but vary using another field variables,
{\it e.g.}, $\gamma_5 \Psi$; or  ${1\over \sqrt{2}} (1-i\gamma_5 ) \Psi$,
what could we obtain?

\end{itemize}

\smallskip

One can go further: I present the Majorana-Ahluwalia
ideas~\cite{DVA2,DVA3}.  We begin with introduction of second-type
4-spinors defined by the formulas:
\begin{equation}
\lambda(p^\mu)\,\equiv
\pmatrix{ \left ( \zeta_\lambda\,\Theta_{[j]}\right
)\,\phi^\ast_{_L}(p^\mu)\cr \phi_{_L}(p^\mu)} \,\,,\quad
\rho(p^\mu)\,\equiv \pmatrix{ \phi_{_R}(p^\mu)\cr \left (
\zeta_\rho\,\Theta_{[j]}\right )^\ast \,\phi^\ast_{_R}(p^\mu)} \,\,\quad
.\label{sp-dva}
\end{equation}
They are not in helicity eigenstates
(see above, Eq. (\ref{wi})), but one can introduce another quantum number,
the chiral helicity, $\eta$ as in ref.~\cite{DVA3}.  Phase
factors $\xi_\lambda$ and $\xi_\rho$ are fixed by the conditions of
self/anti-self $\theta$-conjugacy:
\begin{equation}
S^c_{[1/2]}\lambda
(p^\mu) =\pm \lambda (p^\mu),\quad S^c_{[1/2]}\rho (p^\mu) =\pm \rho
(p^\mu)\quad, \end{equation} for a $j=1/2$ case; and \begin{equation}
\left [\Gamma^5 S^c_{[1]}\right ] \lambda (p^\mu) =
\pm \lambda (p^\mu)\quad,\quad
\left [\Gamma^5 S^c_{[1]}\right ] \rho (p^\mu) =\pm \rho (p^\mu)\quad,
\end{equation}
for a $j=1$ case.
Self/anti-self conjugate spinors do not exist for spin-1 in
the considered model.
Operators of the charge conjugation are defined by formulas
(\ref{sc},\ref{gsc}).  The wave equations for $\lambda$ and $\rho$
spinors, presented by Ahluwalia~\cite{DVA3}, are inconvenient for physical
applications because they have not been written in covariant form:
\begin{equation}\label{eq}
\pmatrix{-\openone &
\zeta_\lambda \,\exp \left ({\bf J}\cdot\mbox{\boldmath $\varphi$}\right )
\Theta_{[j]} \Xi_{[j]} \,\exp \left ({\bf J}\cdot \mbox{\boldmath
$\varphi$}\right )\cr \zeta_\lambda \,\exp \left (-{\bf
J}\cdot\mbox{\boldmath $\varphi$}\right ) \Xi_{[j]}^{-1} \Theta_{[j]}
\,\exp \left (-{\bf J}\cdot \mbox{\boldmath $\varphi$}\right ) & -
\openone\cr} \lambda (p^\mu) =0  \label{weq}
\end{equation}
For the $\rho$ spinors the
equations are obtained by the substitutions $\xi_\lambda \rightarrow
\xi_\rho^\ast$ and $\Theta_{[j]} \Xi_{[j]} \leftrightarrow
\Xi_{[j]}^{-1} \Theta_{[j]}$. Zero-momentum
2-spinors used in Eq. (\ref{sp-dva}) are connected as follows
\begin{equation}
\left [\phi_{_L}^h (\overcirc{p}^\mu) \right ]^* =
\Xi_{[j]} \phi_{_L}^h (\overcirc{p}^\mu)\quad.
\end{equation}
It is this
form of the RB relation which has been used for deriving equations
(\ref{weq}).  Matrices $\Xi_{[j]}$ are defined by the formulas
\begin{equation}
\Xi_{[1/2]} =\pmatrix{e^{i\phi} & 0\cr
0 & e^{-i\phi}\cr},\quad \Xi_{[1]} = \pmatrix{e^{i2\phi} &0 &0\cr
0&1&0\cr 0& 0& e^{-i2\phi}\cr}\quad,
\end{equation}
with $\phi$ being the azimuthal angle associated   with
${\bf p} \rightarrow {\bf 0}$.

Obtained equations appear not to be dynamical equations. One can note
\begin{eqnarray}
\lefteqn{\Theta_{[1/2]}\, \Xi_{[1/2]} = \Xi^{-1}_{[1/2]} \,\Theta_{[1/2]}\,
=\, i\, \frac{\sigma_1 p_2   - \sigma_2 p_1}{\sqrt{(p + p_3)(p - p_3) }}\,
=}\\
&=& \, U_{+} (p^\mu) U_{-} (p^\mu) = -U_+^{-1} (p^\mu) U_-^{-1} (p^\mu)
= - U^{-1}_{\pm} (p^\mu) U_{\pm}
(\tilde p^{\mu}) = U^{-1}_{\pm} (\tilde p^{\mu}) U_{\pm}
(p^\mu)\quad,\nonumber
\end{eqnarray}
where $U_{\pm} (p^\mu)$ are the $2\times 2$
matrices of the unitary transformation  to the helicity
representation~\cite{Berg} and ref.~\cite[p.71]{Novozhilov}:
\begin{equation}
U_+ (p^\mu) \sigma_3 U^{-1}_+ (p^\mu) \,=\,
\frac{(\mbox{\boldmath$\sigma$}\cdot {\bf p})}{p}\quad, \quad U_-^{-1}
(p^\mu) \sigma_3 U_- (p^\mu) \,= \,
-\,\frac{(\mbox{\boldmath$\sigma$}\cdot {\bf p})}{p}\quad;\label{hr1}
\end{equation} $p=\vert {\bf p}\vert =\sqrt{E^2 -m ^2}$ and $\tilde
p^{\mu}$ is the parity-conjugated momentum.  Let us introduce the unitary
matrices
\begin{eqnarray}
{\cal U}_\pm = \pmatrix{ U_\pm (\tilde p^{\mu})
& 0\cr 0 & U_\pm (p^\mu)\cr}\quad,\quad \widetilde {\cal U}_\pm =
\pmatrix{ U_\pm (p^{\mu}) & 0\cr 0 & U_\pm (\tilde
p^\mu)\cr}\quad,\quad\label{Eq}
\end{eqnarray}
which transform to the new
``helicity" representations.  In these new  representations
the equations (\ref{weq}) are presented by:
\begin{eqnarray}
\left [\zeta_\lambda \gamma^5 \gamma^0 - \openone \right ]
\lambda_{_{H}} (p^\mu) \, &=& \,0\quad,\label{equ1}\\
\left
[\zeta_\lambda \gamma^5 \gamma^0 + \openone \right ]
\tilde \lambda_{_{H}} (p^\mu) \, &=& \,0\quad.\label{equ2}
\end{eqnarray}
Analogous results can be obtained in the light-front representation.
{}From the analysis of the general parametrization~\cite{Varshalovich,DVA3}
of 2-spinors  in terms of the polar $\theta$ and the
azimuthal $\phi$
angles associated with the   vector ${\bf p} \rightarrow 0$
(we use the same symbol $\xi_{\pm h}$ for
$\phi_{_L} (\overcirc{p}^\mu)$ and $\phi_{_R} (\overcirc{p}^\mu)$ here):
\begin{eqnarray}
\xi_{+\,1/2} \,&=&\, N e^{i\,\vartheta_1^{_{R,L}}} \pmatrix{\cos\, (\theta/2)
\, e^{-i\phi/2} \cr
\sin \, (\theta /2)\, e^{+i\phi/2} \,} \quad,\\
\xi_{-\,1/2} \,&=&\, N e^{i\,\vartheta^{_{R,L}}_2}
\pmatrix{\sin \, (\theta/2) \, e^{-i\,\phi/2} \cr
- \cos\, (\theta /2)\, e^{+i\phi/2}\,} \quad,
\end{eqnarray}
for spin-$1/2$;
and
\begin{eqnarray}
\xi_{+1} =
N\,e^{i\delta_1^{^{R,L}}}\,\pmatrix{{1\over 2} (1+\cos\theta) e^{-i\phi}\cr
\sqrt{{1\over 2}} \sin\theta\cr
{1\over 2} (1-\cos\theta) e^{+i\phi}\cr}&,&
\xi_{-1} = N\,e^{i\delta_3^{^{R,L}}}\,\pmatrix{{1\over 2} (1-\cos\theta)
e^{-i\phi}\cr
-\sqrt{{1\over 2}} \sin\theta\cr
{1\over 2} (1+\cos\theta) e^{+i\phi}\cr} \\
&&\nonumber\\
\xi_{0} &=&   N\,e^{i\delta_2^{^{R,L}}}\,
\pmatrix{-\sqrt{{1\over 2}}\sin \theta \,e^{-i\phi}\cr
\cos\theta\cr
\sqrt{{1\over 2}}\sin \theta \,e^{+i\phi}\cr}\quad,
\end{eqnarray}
for spin-$1$,
one can find another forms of the RB relation, connecting 2-spinors of
the opposite helicity. They are
\begin{equation}
\left [\phi_{_L}^h
(\overcirc{p}^\mu)\right ]^* = (-1)^{1/2-h} e^{-i(\theta_1 +\theta_2)}
\Theta_{[1/2]} \phi_{_L}^{-h} (\overcirc{p}^\mu)\quad ,\label{rbug12}
\end{equation}
for a $j=1/2$ case;
and
\begin{equation}
\left [\phi_{_L}^h
(\overcirc{p}^\mu)\right ]^* = (-1)^{1-h} e^{-i\delta}
\Theta_{[1]} \phi_{_L}^{-h} (\overcirc{p}^\mu)\quad ,\label{rbug13}
\end{equation}
for a $j=1$ case ($\delta=\delta_1 +\delta_3$ for $h=\pm 1$ and
$\delta=2\delta_2$, for $h=0$). As a result of the use of general
procedure of deriving wave equations~\cite{Ryder,DVA3} we come to the
equations (\ref{i1},\ref{i2}).  Spinors $\Upsilon_{\pm} (p^\mu)$
and ${\cal B}_\pm
(p^\mu)$ are in helicity eigenstates.  They can be presented in the
following form
\begin{eqnarray}
\Upsilon_\pm (p^\mu) = \pmatrix{\pm
i\Theta_{1/2} \left [\phi_{_L}^{\mp 1/2} (p^\mu)\right ]^*\cr
\phi_{_L}^{\pm 1/2} (p^\mu)\cr}\,,\quad {\cal B}_\pm (p^\mu) =
\pmatrix{\mp i\Theta_{1/2} \left [\phi_{_L}^{\mp 1/2} (p^\mu)\right ]^*\cr
\phi_{_L}^{\pm 1/2} (p^\mu)\cr}\,\, ,
\end{eqnarray}
or
\begin{eqnarray}
\widetilde \Upsilon_\pm (p^\mu)=
\pmatrix{\phi_{_R}^{\pm 1/2} (p^\mu)\cr
\mp i\Theta_{1/2} \left [\phi_{_R}^{\mp 1/2} (p^\mu)\right
]^*\cr}\,,\quad
\widetilde {\cal B}_\pm (p^\mu) = \pmatrix{\phi_{_R}^{\pm
1/2}(p^\mu)\cr \pm i\Theta_{1/2} \left [\phi_{_R}^{\mp 1/2} (p^\mu)\right
]^*\cr}\,\, ,
\end{eqnarray}
if $\theta_1 +\theta_2=0$.
Of course, the latter can differ from the former only by
a phase factor provided that we keep the ordinary normalization of
2-spinors.
Moreover, one can note that zero-momentum `Dirac-like  spinors' are
connected with `Majorana-like spinors':
\begin{eqnarray}
\Upsilon_{+1/2}
(\overcirc{p}^\mu) &=& U \lambda^S_\uparrow
(\overcirc{p}^\mu) = -\gamma_5 U \lambda^A_\uparrow
(\overcirc{p}^\mu)\quad,\quad\\
{\cal B}_{+1/2} (\overcirc{p}^\mu) &=& U \lambda^A_\uparrow
(\overcirc{p}^\mu) = -\gamma_5 U \lambda^S_\uparrow
(\overcirc{p}^\mu)\quad,\quad\\
\Upsilon_{-1/2} (\overcirc{p}^\mu) &=&
U \lambda^S_\downarrow (\overcirc{p}^\mu) = -\gamma_5 U \lambda^A_\downarrow
(\overcirc{p}^\mu)\quad,\quad\\
{\cal B}_{-1/2} (\overcirc{p}^\mu) &=& U \lambda^A_\downarrow
(\overcirc{p}^\mu) = -\gamma_5 U \lambda^S_\downarrow
(\overcirc{p}^\mu) \quad.
\end{eqnarray}
The transformation matrix is
\begin{eqnarray}
U = \pmatrix{\Xi^{-1}_{[1/2]} \Theta^{-1}_{[1/2]} & 0\cr
0 & \openone\cr}\quad.
\end{eqnarray}
Other important relations between arbitrary-momentum
`Di\-rac-like' and `Ma\-jo\-ra\-na-like' 4-spinors are:
\begin{eqnarray}
\Upsilon_+ (p^\mu)
&=& \pm\frac{1+\gamma_5}{2} \lambda^{S,A}_\downarrow (p^\mu) +
\frac{1-\gamma_5}{2} \lambda^{S,A}_\uparrow (p^\mu)\quad,\\ \Upsilon_-
(p^\mu) &=& \mp\frac{1+\gamma_5}{2} \lambda^{S,A}_\uparrow (p^\mu) +
\frac{1-\gamma_5}{2} \lambda^{S,A}_\downarrow (p^\mu)\quad,\\ {\cal B}_+
(p^\mu) &=& \mp\frac{1+\gamma_5}{2} \lambda^{S,A}_\downarrow (p^\mu) +
\frac{1-\gamma_5}{2} \lambda^{S,A}_\uparrow (p^\mu)\quad,\\ {\cal B}_-
(p^\mu) &=& \pm\frac{1+\gamma_5}{2} \lambda^{S,A}_\uparrow (p^\mu) +
\frac{1-\gamma_5}{2} \lambda^{S,A}_\downarrow (p^\mu)\quad.
\end{eqnarray}
Analogous relations exist  between $\tilde \Upsilon (p^\mu)$
$\tilde {\cal B} (p^\mu)$ and $\rho^{S,A} (p^\mu)$
spinors.
Finally, treating $\lambda^S$ and $\rho^A$ as positive-energy solutions,
$\lambda^A$ and $\rho^S$ as negative-energy solutions,
the wave equations in the coordinate space are written:
\begin{eqnarray}
i \gamma^\mu \partial_\mu \lambda^S (x) - m \rho^A (x) &=& 0 \quad,
\label{11}\\
i \gamma^\mu \partial_\mu \rho^A (x) - m \lambda^S (x) &=& 0 \quad;
\label{12}
\end{eqnarray}
and
\begin{eqnarray}
i \gamma^\mu \partial_\mu \lambda^A (x) + m \rho^S (x) &=& 0\quad,
\label{13}\\
i \gamma^\mu \partial_\mu \rho^S (x) + m \lambda^A (x) &=& 0\quad.
\label{14}
\end{eqnarray}
As opposed to Eq. (\ref{equ1},\ref{equ2}) they are dynamical equations.
Dynamical parts of the equations
for $\lambda^{S,A}$ (and $\rho^{S,A}$) spinors are connected
with mass parts  of $\rho^{A,S}$ (and $\lambda^{A,S}$)
spinors. Equations (\ref{11}-\ref{14})
can be written in the 8-component form as follows:
\begin{eqnarray}
\left [i \Gamma^\mu \partial_\mu - m\right ] \Psi_{(+)} (x) &=& 0\quad,
\label{psi1}\\
\left [i \Gamma^\mu \partial_\mu + m\right ] \Psi_{(-)} (x) &=& 0\quad,
\label{psi2}
\end{eqnarray}
where
\begin{eqnarray}
\Psi_{(+)} (x) = \pmatrix{\rho^A (x)\cr
\lambda^S (x)\cr}\quad,\quad
\Psi_{(-)} (x) = \pmatrix{\rho^S (x)\cr
\lambda^A (x)\cr}\quad
\end{eqnarray}
are {\it dibispinors}. For purposes of future researches
we define the set of $8\times 8$- component $\Gamma$- and $T$-matrices
\begin{eqnarray}
\Gamma^\mu =\pmatrix{0 & \gamma^\mu\cr
\gamma^\mu & 0\cr}\quad,\quad \Gamma^5 = \pmatrix{\gamma^5 & 0\cr
0 &\gamma^5\cr}\quad,\quad {\L}^5 = \pmatrix{\gamma^5 & 0\cr
0 & -\gamma^5\cr}\quad,
\end{eqnarray}
\begin{eqnarray}
T_{11} = \,\pmatrix{\openone & 0 \cr
0 & -\openone\cr}\quad,\quad
T_{01} = \pmatrix{0& \openone \cr
\openone & 0\cr}\quad,\quad
T_{10} =  \, \pmatrix{0 & \openone\cr
- \openone & 0\cr}\quad.
\end{eqnarray}
The latter  are determined within phase factors ({\it e.g.}, $(-1)^k$,
ref.~\cite{Gelfand}).
The useful commutation relation is
\begin{equation} \L^5
\Gamma^\nu - \Gamma^\nu \L^5 = 0 \quad.\label{crel}
\end{equation}
The Lagrangian is then  given by the formula:
\begin{eqnarray}
{\cal L}^{(1)}
&=& {i\over 2} \left [\,\, \overline\Psi_{(+)}\Gamma^\mu \partial_\mu
\Psi_{(+)} - \partial_\mu \overline\Psi_{(+)}\Gamma^\mu \Psi_{(+)} +
\overline\Psi_{(-)}\Gamma^\mu \partial_\mu \Psi_{(-)} -
\partial_\mu \overline\Psi_{(-)}\Gamma^\mu \Psi_{(-)}\,\,\right ]
\nonumber\\
&-&\, m \left [ \overline
\Psi_{(+)} \Psi_{(+)} - \overline \Psi_{(-)} \Psi_{(-)}   \right
]\quad.\label{Lagr}
\end{eqnarray}
It is useful to note that  the
Lagrangian admits following gradient transformations  of the first
kind for $\lambda^{S,A} (x)$ and $\rho^{S,A} (x)$ spinors:
\begin{eqnarray}
\lambda^\prime (x)
\rightarrow (\cos \alpha -i\gamma^5 \sin\alpha) \lambda
(x)\quad,\label{g10}\\
\overline \lambda^{\,\prime} (x) \rightarrow
\overline \lambda (x) (\cos \alpha - i\gamma^5
\sin\alpha)\quad,\label{g20}\\
\rho^\prime (x) \rightarrow   (\cos \alpha +
i\gamma^5 \sin\alpha) \rho (x) \quad,\label{g30}\\
\overline \rho^{\,\prime} (x) \rightarrow   \overline \rho (x)
(\cos \alpha + i\gamma^5 \sin\alpha)\quad.\label{g40}
\end{eqnarray}
We still note, in general, different gradient transformations of
4-spinors $\lambda$ and $\rho$ are possible.
In terms of field functions $\Psi_{(\pm)} (x)$ the equations
(\ref{g10}-\ref{g40}) are written
\begin{eqnarray}
\Psi^{\,\prime}_{(\pm)}
(x) \rightarrow \left ( \cos \alpha + i \L^5 \sin\alpha \right )
\Psi_{(\pm)} (x)\quad,\label{g1}\\ \overline\Psi_{(\pm)}^{\,\prime} (x)
\rightarrow \overline \Psi_{(\pm)} (x) \left ( \cos \alpha - i \L^5
\sin\alpha \right )\quad.\label{g2}
\end{eqnarray}
Local gradient
transformations are introduced after ``covariantization" of derivatives:
\begin{eqnarray} && \partial_\mu \rightarrow \nabla_\mu = \partial_\mu -
ig \L^5 A_\mu\quad,\\ && A_\mu^\prime (x) \rightarrow A_\mu + {1\over g}
\,\partial_\mu \alpha \quad.  \end{eqnarray} This tells us that the states
described by the spinors of the second type can possess the axial charge.

The second remark:  neither $\lambda^{S,A}$ nor $\rho^{S,A}$ are the
eigenfunctions of the Hamiltonian in the $(1/2,0)\oplus (0,1/2)$
representation space:  \begin{eqnarray} \partial_\mu \gamma^\mu
\lambda_\eta (x) + \wp_{_{\uparrow\downarrow}} m\lambda_{-\eta}
(x)&=&0\quad , \label{cr1}\\ \partial_\mu \gamma^\mu \rho_\eta (x) +
\wp_{_{\uparrow\downarrow}} m\rho_{-\eta} (x)&=&0\quad ,
\label{cr2}
\end{eqnarray}
(the indices $\eta$ should be referred to the chiral helicity introduced in
the Ahluwalia's paper~\cite{DVA3}).
Therefore, the matrix element
$<\lambda^A (0),\,\downarrow \vert \lambda^S (t),
\, \uparrow >$ (and others) have the non-zero values at the time $t$, what
produces speculations on the possibility of oscillations between
self- and anti-self charge conjugate states.
But, the wavelength of these oscillations is very small, of the order of
the de Broglie wavelength. The only effect,
which could be  seen, is the average depletion of the flux composed from
the pure, {\it e.g.}, $\lambda^S$ states.

We should point out relations with other papers, {\it e.g.},
ref.~\cite{Ziino,Barut}.
Authors presented the  model based on the following principles: a $j=1/2$
particle is described by $\Psi_f (x)$;  its antiparticle, by
$\Psi_{\overline f} (x)$. These wave functions
satisfy different equations:
\begin{eqnarray}
\left [i\gamma^\mu \partial_\mu -m \right ] \Psi_f (x)&=&0\quad,\\
\left [i\gamma^\mu \partial_\mu +m \right ] \Psi_{\overline f } (x)&=&0\quad.
\end{eqnarray}
In the framework of their model
the role of the charge conjugation matrix is played by $\gamma^5$ matrix.
As a matter of fact this model is recreation of the ideas of
Belinfante, Pauli~\cite{Belin} and, particularly, of
Professor  M. Markov ~\cite{Markov}. Barut-Ziino asymptotically
chiral massive fields:
\begin{equation}
\psi_f^{ch} = \frac{\psi_f - \psi_{\bar f}}{\sqrt{2}}\quad,\quad
\psi_{\overline f}^{ch} = \frac{\psi_f +\psi_{\bar f}}{\sqrt{2}}
\end{equation}
have close relations with $\lambda^A , \rho^A \sim \psi^{ch}_f$,
$\lambda^S , \rho^S \sim \psi^{ch}_{\overline f}$.
In this connection Barut and Ziino noted on
needed modifications of our understanding  of  the concept of the
quantization space~\cite{Barut}: ``Such a Fock space should have
the manifestly
covariant structure
\begin{equation}
{\cal F} \equiv {\cal F}^0 \otimes
S_{in}\quad,
\end{equation}
where ${\cal F}^0$ is an ordinary Fock space
for one {\it indistinct} type of positive- and negative-energy identical
spin-${1\over 2}$ particles (without regard to the proper-mass sign) and
$S_{in}$ is a two-dimensional internal space spanned by the proper-mass
eigenstates $\vert +m >$, $\vert -m >$ thus {\it doubling} ${\cal F}^0$.
This allows [the Dirac-like fields] $\psi_f$ and $\psi_{\overline f}$
to be {\it mixed} if a rotation is
performed in $S_{in}$." Moreover, the authors of~\cite{Ziino,Barut}
noted, indeed, that the effect of `parity violation' can be explained
in the framework of parity-invariant theory. While the states are not,
in general, eigenstates of Parity operator the viewpoint of Barut and
Ziino is preferable because it lifts the crucial contradiction with
relativity:  there are no any reasons Nature to consider left- and
right-handed frames in unsymmetrical fashion. As a matter of fact,
possibility of such constructs (the former case is called {\it doubling})
has been discovered by Wigner~\cite{Wigner2}, who enumerated the
irreducible projective representations of the full Poincar\`e group
(including reflections). The constructs given  by Professor D. V.
Ahluwalia {\it  et al.} present themselves explicit examples of the
theories of such a type. Neutrino and its antineutrino (of the same
chirality) could coincide as a particular case of this model.

I would like to finish my talk by the question which I ask you
in the beginning, the question of the claimed longitudity
of  the antisymmetric tensor field. Authors of previous works
treated both ``gauge"-invariant Lagrangians
and conformal-invariant Lagrangians. Nevertheless, I can not accept so
obvious violation of the Correspondence Principle.  The contradiction
appears to me to be related with the problem of acausal non-plane wave
solutions~\cite{DVA00} in the first-order equations of the form (4),
ref.~\cite{Recami}, and (4.21,4.22) in~[34b,p.B888].  In the classical
electromagnetic field theory we know that physical field variables are the
strengths ${\bf E}$ and ${\bf B}$.  Potentials are used only as a
convenient way to calculate the former.  Aharonov and Bohm~\cite{Aharonov}
told us that this is not the case in the quantum theory. Potentials
appear to have physical significance. However, attempts to construct the
quantized electromagnetic theory based on the use of physical variables
(in fact, on the $(1,0)\oplus (0,1)$ representation of the Poincar\`e
group, {\it i.e.}, on the first principles) have also certain reasons.
Moreover, in ref.~\cite{Weinberg2} Professor S.  Weinberg  proved that in
the quantized theory the 4-vector potential $A_\mu$ is not a 4-vector (!)
at all.

One must begin with the general form of the field operator:
\begin{equation}
F^{\mu\nu} (x) \,=\,
\sum_h \int \frac{d^3 {\bf p}}{2E_p (2\pi)^3} \left [
F^{\mu\nu}_{h\, (+)} ({\bf p})\, a_h ({\bf p})\, e^{-ip\cdot x} +
F^{\mu\nu}_{h\, (-)} ({\bf p}) \,b_h^\dagger ({\bf p})\, e^{+ip\cdot x}
\right ]\, .
\end{equation}
Of course, it is very important
question, what we understand under $F^{\mu\nu}_{h\, (+)} ({\bf p})$ and
$F^{\mu\nu}_{h\,(-)} ({\bf p})$. One can not forget about the possibility
of the use of components of the dual tensor.

The Lagrangian is
\begin{equation}
{\cal L} = {1\over 4} (\partial_\mu
F_{\nu\alpha})(\partial^\mu F^{\nu\alpha}) - {1\over 2} (\partial_\mu
F^{\mu\alpha})(\partial^\nu F_{\nu\alpha}) - {1\over 2} (\partial_\mu
F_{\nu\alpha})(\partial^\nu F^{\mu\alpha}) +{1\over 4} F_{\mu\nu}
F^{\mu\nu} \quad.\label{Lagran}
\end{equation}
The massless limit ($m\rightarrow 0$) of this Lagrangian is compatible
with conformal invariance (and with the ``gauge" invariance within the
generalized Lorentz condition, see~\cite{Hayashi,DVO01}).
The Lagrange-Euler equation is then written
\begin{equation}
{1\over 2} ({\,\lower0.9pt\vbox{\hrule
\hbox{\vrule height 0.2 cm \hskip 0.2 cm \vrule height
0.2cm}\hrule}\,} +m^2)  F_{\mu\nu} +
(\partial_{\mu}F_{\alpha\nu}^{\quad,\alpha} -
\partial_{\nu}F_{\alpha\mu}^{\quad,\alpha}) = 0 \quad,
\end{equation}
where ${\,\lower0.9pt\vbox{\hrule \hbox{\vrule height 0.2 cm
\hskip 0.2 cm
\vrule height 0.2 cm}\hrule}\,}
=- \partial_{\alpha}\partial^{\alpha}$.
Similar equations follow from  the Proca
theory (one should take into account the Klein-Gordon equation
in the final expression):
\begin{eqnarray}
&&\partial_\alpha F^{\alpha\mu} + m^2 A^\mu = 0 \quad, \\
&&F^{\mu\nu} = \partial^\mu A^\nu - \partial^\nu A^\mu \quad; \label{Proca}
\end{eqnarray}
and from the Weinberg's $2(2j+1)$- component theory, provided that
${\bf E}$ and ${\bf B}$ are treated to be physical
variables. They form the $j=1$ Weinberg's wave function.

The variation procedure, ref.~\cite{Bogoliubov,Corson}, for rotation:
\begin{equation}
x^{\mu^\prime} = x^\mu + \omega^{\mu\nu} x_\nu
\end{equation}
leads to
\begin{equation}
\delta F^{\alpha\beta} = {1\over 2} \omega^{\kappa\tau}
{\cal T}_{\kappa\tau}^{\alpha\beta,\mu\nu} F_{\mu\nu}\quad.
\end{equation}
Generators of infinitesimal transformations are defined as
\begin{eqnarray}
\lefteqn{{\cal T}_{\kappa\tau}^{\alpha\beta,\mu\nu} \,=\,
{1\over 2} g^{\alpha\mu} (\delta_\kappa^\beta \,\delta_\tau^\nu \,-\,
\delta_\tau^\beta\,\delta_\kappa^\nu) \,+\,{1\over 2} g^{\beta\mu}
(\delta_\kappa^\nu\delta_\tau^\alpha   \,-\,
\delta_\tau^\nu\, \delta_\kappa^\alpha) +\nonumber}\\
&+&\,
{1\over 2} g^{\alpha\nu} (\delta_\kappa^\mu \, \delta_\tau^\beta \,-\,
\delta_\tau^\mu \,\delta_\kappa^\beta) \,+\, {1\over 2}
g^{\beta\nu} (\delta_\kappa^\alpha \,\delta_\tau^\mu \,-\,
\delta_\tau^\alpha \, \delta_\kappa^\mu)\quad.
\end{eqnarray}
The classical  formula for the Pauli-Lyuban'sky operator
is obtained immediately (${\bf n} \vert\vert {\bf p}$):
\begin{eqnarray}
(W\cdot n) = m \epsilon^{ijk} n^i \int d^3 {\bf x} \left [ F_{0}^{\,\, j}
(\partial^\mu F_{\mu}^{\,\, k})
+ F^{\mu k} (\partial_0 F_{\mu}^{\,\, j} +\partial_\mu F^{j}_{\,\,0}
+\partial^j F_{0\mu} ) \right ]\quad.\label{PL}
\end{eqnarray}
Let us remind that massless limit of this theory, ref.~\cite{Weinberg},
is well-defined. One can
substitute the field operator in this expression and we can learn,
under what  constraints the Pauli-Lyuban'sky operator is
equal to zero, why in the previous works the conclusion has been done that
massless antisymmetric tensor field appears to be longitudinal (?), and
what corresponds the longitudinal solution to.

\acknowledgements

I greatly appreciate   many useful advises of   Profs. D. V. Ahluwalia,
A. E. Chubykalo, M.~W. Evans, I.~G. Kaplan, A.~F. Pashkov and Yu. F.
Smirnov.  I   thank participants of the Escuela Latino Americana de
F\'{\i}sica (ELAF'95) and the Int.  Conf.  on the Theory of the Electron
(ICTE'95) for interest in my work.

I am grateful to Zacatecas University for  professorship.

This work has been partially supported by Mexican Sistema Nacional de
Investigadores and Programa de Apoyo a la Carrera Docente.

\end{document}